\documentclass[12pt,chicago,iop]{emulateapj}

\usepackage{natbib} 
\usepackage{bm}
\usepackage{times}
\usepackage{graphicx}
\usepackage{color}
\usepackage{amssymb}
\usepackage{fancyvrb}
\usepackage{verbatim}

\renewcommand{\cite}{\citep}
\newcommand{\HI}{H$\,${\sc i}}
\newcommand{\rHI}{\rm HI}

\begin{document}
\VerbatimFootnotes

\title{Measurement of 21\,cm brightness fluctuations at $\lowercase{z} \sim 0.8$ in
cross-correlation
\shorttitle{21\,cm brightness fluctuations}}

\author{
K.~W.~Masui\altaffilmark{1,2},
E.~R.~Switzer\altaffilmark{1,3},
N.~Banavar\altaffilmark{4},
K.~Bandura\altaffilmark{5},
C.~Blake\altaffilmark{6},
L.-M.~Calin\altaffilmark{1},
T.-C.~Chang\altaffilmark{7},
X.~Chen\altaffilmark{8,9},
Y.-C.~Li\altaffilmark{8},
Y.-W.~Liao\altaffilmark{7},
A.~Natarajan\altaffilmark{10},
U.-L.~Pen\altaffilmark{1},
J.~B.~Peterson\altaffilmark{10},
J.~R.~Shaw\altaffilmark{1},
T.~C.~Voytek\altaffilmark{10}
}

%

\altaffiltext{1}{Canadian Institute for Theoretical Astrophysics,
University of Toronto, 60 St. George St., Toronto, Ontario, M5S 3H8}
\altaffiltext{2}{Department of Physics, University of Toronto,
60 St. George St., Toronto, Ontario, M5S 1A7, Canada}
\altaffiltext{3}{Kavli Institute for Cosmological Physics,
University of Chicago,
5640 South Ellis Avenue, Chicago, IL 60637, USA}
\altaffiltext{4}{Department of Astronomy \& Astrophysics, University of Toronto,
50 St. George St., Toronto, Ontario, M5S 3H4, Canada}
\altaffiltext{5}{Department of Physics, McGill University, 3600 Rue University,
Montreal, Quebec, H3A 2T8, Canada}
\altaffiltext{6}{Centre for Astrophysics \& Supercomputing, Swinburne University
of Technology, P.O. Box 218, Hawthorn, VIC 3122, Australia}
\altaffiltext{7}{Academia Sinica Institute of Astronomy and Astrophysics, PO
Box 23-141, Taipei, 10617, Taiwan}
\altaffiltext{8}{National Astronomical Observatories,
Chinese Academy of Science, 20A Datun Road, Beijing 100012, China}
\altaffiltext{9}{Center of High Energy Physics, Peking University, Beijing, 100871, China}
\altaffiltext{10}{McWilliams Center for Cosmology, Carnegie Mellon University,
Department of Physics, 5000 Forbes Ave., Pittsburgh PA 15213, USA}

\shortauthors{Masui, Switzer, et. al.}


\begin{abstract}
In this letter, 21\,cm intensity maps acquired at the Green
Bank Telescope are cross-correlated with large-scale structure traced by
galaxies in the WiggleZ Dark Energy Survey.  The data span the redshift range
$0.6 < z < 1$ over two fields totaling $\sim41$~deg.~sq. and $190$~hr of
radio integration time.  The cross-correlation constrains $\Omega_{\rHI}
b_{\rHI}r=[0.43\pm0.07({\rm stat.})\pm0.04({\rm sys.})]\times
10^{-3}$, where $\Omega_{\rHI}$ is the neutral hydrogen (\HI) fraction, $r$
is the galaxy--hydrogen correlation coefficient, and $b_{\rHI}$ is the \HI\
bias parameter.  This is the most precise constraint on neutral hydrogen
density fluctuations in a challenging redshift range.  Our measurement improves
the previous 21\,cm cross-correlation at $z\sim0.8$ both in its precision and
in the range of scales probed.
\end{abstract}


\maketitle

\section{Introduction}

Measurements of neutral hydrogen are essential to our understanding of the
universe.  Following cosmological reionization at $z\sim6$, the
majority of hydrogen outside of galaxies is ionized.  Within galaxies, it must
pass through its neutral phase (\HI) as it cools and collapses to form stars.
The quantity and distribution of neutral hydrogen is therefore intimately
connected with the evolution of stars and galaxies, and observations of neutral
hydrogen can give insight into these processes.

Above redshift $z=2.2$, the Ly-$\alpha$ line redshifts into optical
wavelengths and \HI\ can be observed, typically in absorption against distant
quasars \cite{2009ApJ...696.1543P}.
Below redshift $z=0.1$, \HI\ has been studied
using 21\,cm emission from its hyperfine splitting \cite{2005MNRAS.359L..30Z,
2010ApJ...723.1359M}.  There, the abundance and large-scale distribution of
neutral hydrogen are inferred from large catalogs of discrete galactic
emitters.
Between $z=0.1$ and $z=2.2$ there are fewer constraints on neutral
hydrogen, and those that do exist \cite{2011ApJ...732...35M,
2007MNRAS.376.1357L, 2006ApJ...636..610R} have large uncertainties.

While the 21\,cm line is too faint to observe individual galaxies in this
redshift range, one can nonetheless pursue three-dimensional (3D) intensity mapping
\cite{2008PhRvL.100i1303C, 2008PhRvL.100p1301L, 2012A&A...540A.129A,
2008PhRvD..78b3529M, 2010ApJ...721..164S, 2012ApJ...752...80M}.  Instead of
cataloging many individual galaxies, one can study the large-scale structure
(LSS) directly by detecting the aggregate emission from many galaxies that
occupy large $\sim1000\,{\rm Mpc}^3$  voxels.  The use of such large voxels
allows telescopes such as the Green Bank Telescope (GBT) to reach $z\sim1$,
conducting a rapid survey of a large volume.

Aside from being used to measure the hydrogen content of galaxies,
intensity mapping promises to be an efficient way to study the large-scale
structure of the Universe. In particular, the method could be used to measure
the baryon acoustic oscillations to high accuracy and constrain dark energy
\citep{2008PhRvL.100i1303C}. However, intensity mapping is a new technique 
which is still being pioneered. Ongoing observational efforts such as the one
presented here are essential for developing this technique as a powerful probe
of cosmology.

Synchrotron foregrounds are the primary challenge to this method, because they
are three orders of magnitude brighter than the 21\,cm signal.
However, the physical process of synchrotron emission is known to produce
spectrally smooth radiation \cite{2003MNRAS.346..871O, 2010ApJ...721..164S}.
If the calibration, spectral response and beam width of the instrument are
well-controlled and characterized, the subtraction of foregrounds should be
possible because the foregrounds have fewer degrees of freedom than the
cosmological signal.  We find that this allows the foregrounds to be cleaned to
the level of the expected signal.  The auto-correlation of intensity maps is
biased by residual foregrounds, and minimizing and constraining these residuals
is an active area of work.  However, because residual foregrounds should be
uncorrelated with the cosmological signal, they only boost the
noise in a cross-correlation with existing surveys. This makes the
cross-correlation a robust indication of neutral hydrogen density fluctuations 
in the 21\,cm intensity maps \cite{2010Natur.466..463C, 2012A&A...539L...5V}.

The first detection of the cross-correlation between LSS and 21\,cm intensity
maps at $z\sim1$ was reported in \citet{2010Natur.466..463C}, based on data
from GBT and the DEEP2 galaxy survey.  Here we improve on these measurements by
cross correlating new intensity mapping data with the WiggleZ Dark Energy
Survey \cite{2010MNRAS.401.1429D}.  Our measurement improves on the statistical
precision and range of scales of the previous result, which was based on
15\,hr of GBT integration time over 2~deg.~sq.

Throughout, we use cosmological parameters from \citet{2009ApJS..180..330K}, in
accord with \citet{2011MNRAS.415.2876B}.

\section{Observations}

The observations presented here were conducted with the $680$--$920$~MHz
prime-focus receiver at the GBT. The unblocked aperture
of GBT's $100$~m offset paraboloid design results in well-controlled sidelobes
and ground spill, advantageous to minimizing radio-frequency contamination and
overall system temperature ($\sim25$~K). The receiver is sampled from
$700$~MHz ($z=1$) to $900$~MHz ($z=0.58$) by the Green Bank Ultimate Pulsar
Processing Instrument (GUPPI) pulsar back-end systems
\citep{2008SPIE.7019E..45D}.

The data used in this analysis were collected between 2011 February and November
as part of a 400\,hr allocation over four fields. This allocation was
specifically to corroborate previous cross-correlation measurements
\citep{2010Natur.466..463C} over a larger survey area, and to search for
auto-power of diffuse $21$~cm emission.  The analysis here is based on a
105\,hr integration of a $4.5^\circ\times2.4^\circ$ ``$15$~hr deep field"
centered at $14^\mathrm{h}31^\mathrm{m}28.5^\mathrm{s}$ right ascension, 
$2^\circ0'$ declination and an
84\,hr integration on a $7.0^\circ\times4.3^\circ$ ``$1$~hr shallow" field
centered at $0^\mathrm{h}52^\mathrm{m}0^\mathrm{s}$ right ascension, 
$2^\circ9'$ declination.  The beam
FWHM at $700$~MHz is $0.314^\circ$ and at $900$~MHz it is $0.25^\circ$.  At
band-center, the beam width corresponds to a comoving length of
$9.6\,h^{-1}$Mpc.  Both fields have nearly complete angular overlap and good
redshift coverage with WiggleZ. 

Our observing strategy consists of sets of azimuthal scans at constant
elevation to control ground spill.  We start the set at the low right ascension
(right hand) side of the field and allow the region to drift through. We then
re-point the telescope to the right side of the field and repeat the process.
For the 15\,hr field, this set of scans consists of 8 one-minute scans each
with a stroke of $4^\circ$. For the 1\,hr field, a set of scans consists of 10
two-minute scans, each $8^\circ$ in length. Note that since we observe over a
range of local sidereal times, our scan directions cover a range of angles with
respect to the sky. This range of crossing angles makes the noise more
isotropic, and allows us to ignore the directional dependence of the noise 
in the 3D power spectrum. 
The survey regions have most coverage in the middle due to the largest number
of intersecting scans.  Observations were
conducted at night to minimize radio-frequency interference (RFI). 

The optical data are part of the WiggleZ Dark Energy Survey
\citep{2010MNRAS.401.1429D}, a large-scale spectroscopic survey of
emission-line galaxies selected from UV and optical imaging. It spans
redshifts $0.2 < z < 1.0$ across $1000$~sq.~deg. The selection function
\citep{2010MNRAS.406..803B} has angular dependence determined primarily by the
UV selection, and redshift coverage which favors the $z=0.6$ end of the radio
band. The galaxies are binned into volumes with the same pixelization as the
radio maps and divided by the selection function, so that we consider the
cross-power with respect to optical over-density.

\section{Analysis}

Here we describe our analysis pipeline, which converts the raw data into 3D
intensity maps, then correlates these maps with the WiggleZ
galaxies.\footnote{Our analysis software is publicly available at {\tt
https://github.com/kiyo-masui/analysis\_IM}}

\subsection{From data to maps}
\label{ss:datatomaps}

The first stage of our data analysis is a rough cut to mitigate contamination
by terrestrial sources of RFI. Our data natively have fine spectral resolution
with 4096 channels across 200\,MHz of bandwidth.  This facilitates the
identification and flagging of RFI.  In each scan, individual channels are
flagged based on their variance.  Any RFI not sufficiently prominent to be
flagged in this stage is detected as increased noise later in the pipeline and
subsequently down-weighted during map-making.  Additional RFI is detected as
frequency-frequency covariance in the foreground cleaning and subtracted in the
map domain.  While RFI is prominent in the raw data, after these steps, it was
not found to be the primary limitation of our analysis.

In addition to RFI, we also eliminate channels within 6\,MHz of the band edges
(where aliasing is a concern) and channels in the 800\,MHz receiver's two
resonances at roughly 798\,MHz and 817\,MHz.  Before mapping, the data are
re-binned to 0.78\,MHz-wide bands (corresponding to roughly $3.8\,h^{-1}$Mpc at
band-center). 

For a time-transfer calibration standard, we inject power from a noise diode
into the antenna. The noise diode raises the system temperature by roughly 2\,K
and we switch it at 16\,Hz so that the noise power can be cleanly isolated.
Calibration is performed by first dividing by the noise diode power (averaged
over a scan) in each channel, and then converting to flux using dedicated
observations of 3C286 and 3C48. The gain for $X$ and $Y$ polarizations may
differentially drift and so these are calibrated independently. 
Our absolute calibration uncertainty is dominated by the calibration
of the reference flux scale ($5\%$, \citet{1969ApJ...157....1K}), measurements
of the calibration sources with respect to this reference ($5\%$, see also
\citet{2012MNRAS.423L..30S}), and uncertainty of our measurement of these fluxes
($5\%$). Receiver nonlinearity, uncertainty in the beam shape and variations in
the diffuse galactic emission in the on- and off-source measurements are
estimated to contribute of order $1\%$ each. These are all assumed to be
uncorrelated errors and give $9\%$ total calibration systematic error.

\begin{figure*}[htb]
\includegraphics[scale=0.72]{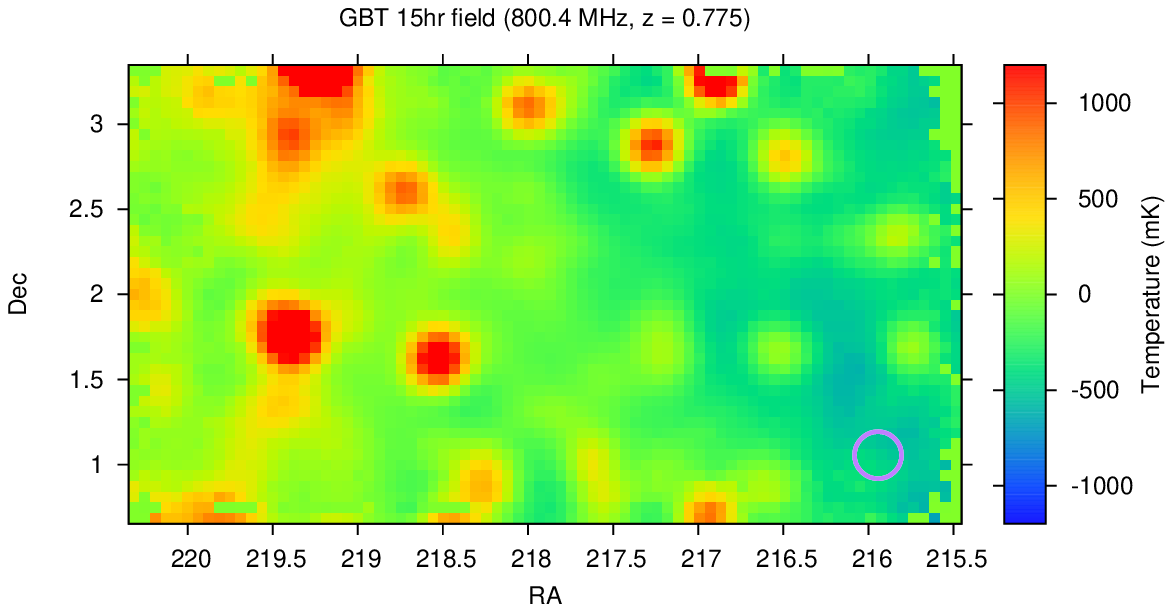}
\includegraphics[scale=0.72]{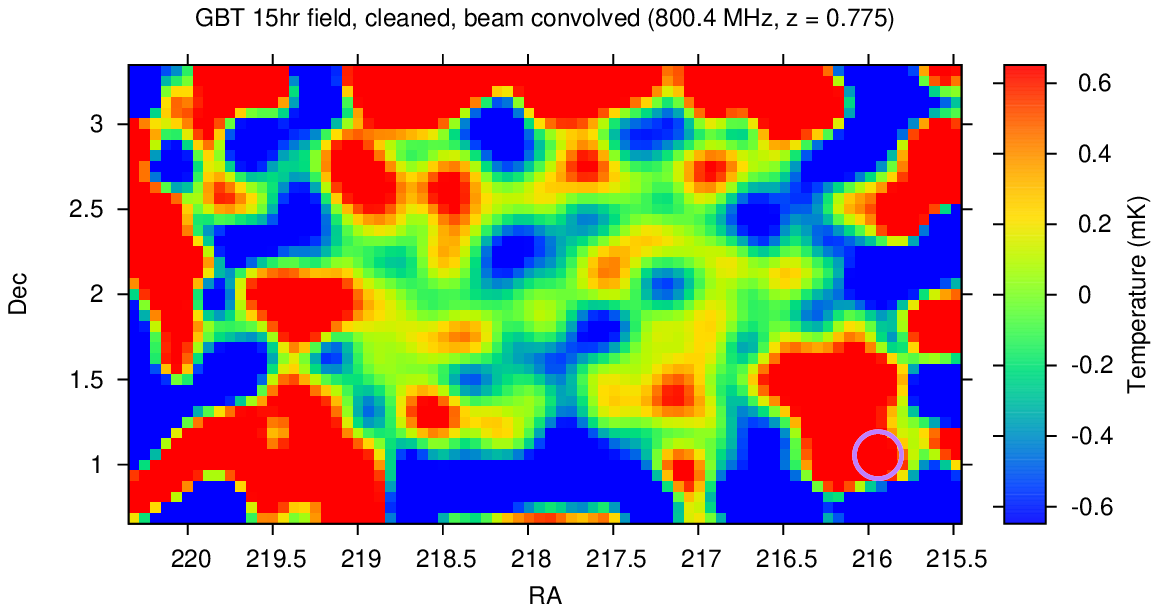}
\caption{\label{f:map15}\label{f:mapclean15}
Maps of the GBT 15\,hr field at approximately the band-center. The purple
circle is the FWHM of the GBT beam, and the color range saturates in some
places in each map.  \emph{Left:} The raw map as produced by the map-maker.  It is
dominated by synchrotron emission from both extragalactic point sources and
smoother emission from the galaxy.  \emph{Right:} The raw map with 20
foreground modes removed per line of sight relative to $256$ spectral bins, as
described in Sec.~\ref{ss:maptopower}.  The map edges have visibly higher noise
or missing data due to the sparsity of scanning coverage.
The cleaned map is dominated by thermal noise,
and we have convolved by GBT's beam shape to
bring out the noise on relevant scales.}
\end{figure*}

Gridding the data from the time ordered data to a map is done in two stages.
We follow cosmic microwave background (CMB) map-making conventions as described in
\citet{1997ApJ...480L..87T}.  The map maker treats the noise to be 
uncorrelated except for deweighting the mean and slope along the time axis for
each scan.
Each frequency channel is treated independently.  In the first round of
map-making, the noise is estimated from the variance of the scan.  This is
inaccurate because the foregrounds dominate the noise.  This
yields a sub-optimal map which nonetheless has high a signal-to-noise ratio on
the foregrounds.  This map is used to estimate the expected foreground signal
in the time ordered data and to subtract this expected signal, leaving time
ordered data which are dominated by noise.  After flagging anomalous data
points at the 4$\sigma$ level, we re-estimate the noise and use this estimate
for a second round of map-making, yielding a map which is much closer to
optimal.  In reality, it is a bad assumption that the noise is uncorrelated.
We have observed correlations at finite time lag and between separate frequency 
channels
in our data.  Exploiting these correlations to improve the optimality of our
maps is an area of active research.  For all map-making, we use square pixels
with widths of 0.0627$^\circ$, which corresponds to a quarter of the beam's
FWHM at the high frequency edge of our band.  Fig.~\ref{f:map15} shows the
15\,hr field map.

In addition to the observed maps, we develop signal-only simulations based on
Gaussian realizations of the non-linear, redshift-space power spectrum using the
empirical-NL model described by \citet{2011MNRAS.415.2876B}.  

\subsection{From maps to power spectra}
\label{ss:maptopower}

The approach to 21\,cm foreground subtraction in literature has been dominated
by the notion of fitting and subtracting smooth, orthogonal polynomials along
each line of sight.  This is motivated by the eigenvectors of smooth
synchrotron foregrounds \citep{2011PhRvD..83j3006L, 2012MNRAS.419.3491L}. In
practice, instrumental factors such as the spectral calibration (and its
stability) and polarization response translate into foregrounds that have more
complex structure. One way to quantify this structure is to use the map itself
to build the foreground model. To do this, we find the frequency-frequency
covariance across the sample of angular pixels in the map, using a noise
inverse weight.  We then find the principal components along the frequency
direction, order these by their singular value, and subtract a fixed number of
modes of the largest covariance from each line of sight. Because the
foregrounds dominate the real map, they also dominate the largest modes of the
covariance.  

There is an optimum in the number of foreground modes to remove.  For too few
modes, the errors are large due to residual foreground variance.  For too many
modes, 21\,cm signal is lost, and so after compensating based on simulated
signal loss (see below), the errors increase modestly. We find that removing 20
modes in both the 15\,hr and 1\,hr field maximizes the signal.
Fig.~\ref{f:mapclean15} shows the foreground-cleaned 15\,hr field map.

We estimate the cross-power spectrum using the inverse noise variance of the
maps and the WiggleZ selection function as the weight for the radio and optical
survey data, respectively. The variance is estimated in the mapping step
and represents noise and survey coverage.  The foreground cleaning process also
removes some 21\,cm signal.  We compensate for signal loss using a transfer
function based on 300 simulations where we add signal simulations to the
observed maps (which are dominated by foregrounds), clean the combination, and
find the cross-power with the input simulation.  Because the foreground
subtraction is anisotropic in $k_\perp$ and $k_\parallel$, we estimate and
apply this transfer function in 2D. The GBT beam acts strictly in $k_\perp$,
and again we develop a 2D beam transfer function using signal simulations with
the beam.

The foreground filter is built from the real map which has a limited number of
independent angular elements.  This causes the transfer function to have
components in both the
angular and frequency direction \citep{Nityananda10}, with the angular part
dominating.
This is accounted for in
our transfer function.  Subtleties of the cleaning method will be described in
a future methods paper. 

\begin{figure}[htb]
\includegraphics[scale=0.72]{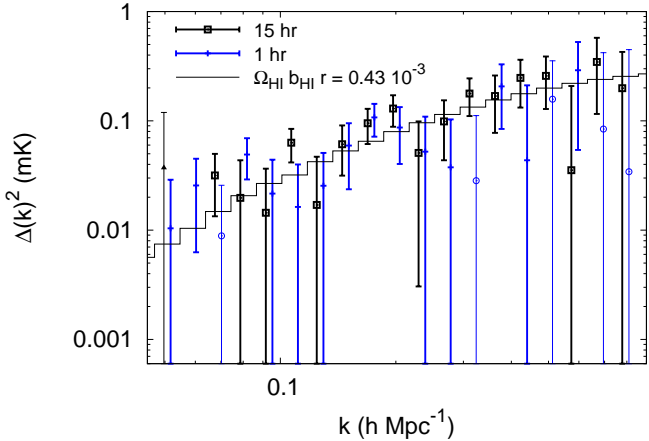}
\caption{\label{f:corrcomb} Cross-power between the 15\,hr and 1\,hr GBT fields and WiggleZ. 
Negative points are shown with reversed sign and a thin line. The solid line is 
the mean of simulations based on the empirical-NL model of \citet{2011MNRAS.415.2876B} 
processed by the same pipeline.}
\end{figure}

We estimate the errors and their covariance in our cross-power spectrum by
calculating the cross-power of the cleaned GBT maps with 100 random catalogs
drawn from the WiggleZ selection function \citep{2010MNRAS.406..803B}.  The mean of
these cross powers is consistent with zero, as expected.  The variance accounts
for shot noise in the galaxy catalog and variance in the radio map either from
real signal (sample variance), residual foregrounds or noise.  Estimating the
errors in this way requires many independent modes to enter each spectral
cross-power bin. This fails at the lowest $k$ values and so these scales are
discarded.  In going from the two-dimensional power to the 1D powers presented
here, we weight each 2D $k$-cell by the inverse variance of the 2D cross-power
across the set of mock galaxy catalogs. The 2D to 1D binning weight is
multiplied by the square of the beam and foreground cleaning transfer
functions.  Fig.~\ref{f:corrcomb} shows the resulting galaxy-\HI\ cross-power
spectra.

\section{Results and discussion}
\label{sec:results}

To relate the measured spectra with theory, we start with the mean 21\,cm
emission brightness temperature \cite{2010Natur.466..463C},
\begin{equation}
T_b=0.29\frac{\Omega_{\rm {HI}}}{10^{-3}}\left(\frac{\Omega_m+
(1+z)^{-3}\Omega_\Lambda}{0.37}\right)^{-\frac{1}{2}}\left(\frac{1+z}{1.8}
\right)^{\frac{1}{2}}~{\rm mK}.
\end{equation}
Here $\Omega_{\rm{HI}}$ is the comoving \HI\ density (in units of today's
critical density), and $\Omega_m$ and $\Omega_\Lambda$ are evaluated at the
present epoch.  We observe the brightness contrast, $\delta T = T_b
\delta_{\rm{HI}}$, from fluctuations in the local \HI\ over-density
$\delta_{\rm{HI}}$.  On large scales, it is assumed that neutral hydrogen and
optically-selected galaxies are biased tracers of the dark matter, so that
$\delta_{\rm{HI}} = b_{\rm{HI}} \delta$, and $\delta_{\rm{opt}} = b_{\rm{opt}}
\delta$.  In practice, both tracers may contain a stochastic component, so we
include a galaxy-\HI\ correlation coefficient $r$. This quantity is
scale-dependent because of the $k$-dependent ratio of shot noise to
large-scale structure, but should approach unity on large scales.  The
cross-power spectrum is then given by $P_{\rm{HI},\rm{opt}}(k)=T_b
b_{\rm{HI}}b_{\rm{opt}}rP_{\delta\delta}(k)$ where $P_{\delta\delta}(k)$
is the matter power spectrum.

The large-scale matter power spectrum is well-known from CMB
measurements \cite{2011ApJS..192...18K} and the bias of the optical
galaxy population is measured to be $b_{\rm{opt}}^2=1.48\pm0.08$ at the
central redshift of our survey \cite{2011MNRAS.415.2876B}. Simulations
including nonlinear scales (as in Sec.~\ref{ss:datatomaps}) are run through the
same pipeline as the data. We fit the unknown prefactor $\Omega_{\rHI}b_{\rHI}r$
of the theory to the measured cross-powers shown in
Fig.~\ref{f:corrcomb}, and determine 
$\Omega_{\rHI}b_{\rHI}r=[0.44\pm0.10({\rm
stat.})\pm0.04({\rm sys.})]\times
10^{-3}$ for the 15\,hr
field data, and 
$\Omega_{\rHI}b_{\rHI}r=[0.41\pm0.11({\rm
stat.})\pm0.04({\rm sys.})]\times10^{-3}$
for the 1\,hr field data.  The systematic
term represents the $9\%$ absolute calibration uncertainty from
Sec.~\ref{ss:datatomaps}. It does not include current uncertainties in the
cosmological parameters or in the WiggleZ bias, but these are
sub-dominant.
Combining the two fields yields 
$\Omega_{\rHI}b_{\rHI}r=[0.43\pm0.07({\rm stat.})\pm0.04({\rm
sys.})]\times10^{-3}$.  These fits are based on
the range $0.075\,h{\rm Mpc}^{-1}<k<0.3\,h{\rm Mpc}^{-1}$ over which we
believe that errors are well-estimated (failing toward larger scales where
there are too few $k$ modes in the volume) and under the assumption that
nonlinearities and the beam/pixelization (failing toward smaller scales) are
well-understood. A less conservative approach is to fit for $0.05\,h{\rm
Mpc}^{-1}<k<0.8\,h{\rm Mpc}^{-1}$ where the beam, model of nonlinearity and
error estimates are less robust, but which shows the full statistical power of
the measurement, at $7.4\sigma$ combined. Here, 
$\Omega_{\rHI}b_{\rHI}r=[0.40\pm0.05({\rm stat.})\pm0.04({\rm sys.})]\times10^{-3}$
for the
combined, $\Omega_{\rHI}b_{\rHI}r=[0.46\pm0.08]\times10^{-3}$ for
the 15\,hr field and $\Omega_{\rHI}b_{\rHI}r=[0.34\pm0.07]\times10^{-3}$
for the 1\,hr field.

To compare to the result in \citet{2010Natur.466..463C}, 
$\Omega_{\rHI}b_{\rm rel}r=[0.55\pm0.15({\rm stat.})]\times10^{-3}$,
we must multiply
their relative bias (between the GBT intensity map and DEEP2) by the DEEP2 bias
$b=1.2$ \citep{2004ApJ...609..525C} to obtain an expression with respect to
$b_{\rHI}$. This becomes $\Omega_{\rHI}b_{\rHI}r=[0.66\pm0.18({\rm
stat.})]\times10^{-3}$, and is consistent with our result.

The absolute abundance and clustering of \HI\ 
are of great interest in studies of galaxy and star formation.  
Our measurement is an integral constraint on the \HI\ luminosity function,
which can be directly compared to simulations.
The quantity $\Omega_{\rHI}b_{\rm
HI}$ also determines the amplitude of 21\,cm temperature fluctuations. This is
required for forecasts of the sensitivity of future 21\,cm intensity mapping
experiments.
Since $r<1$ we have put a lower limit on $\Omega_{\rHI}b_{\rHI}$. 

To determine $\Omega_{\rHI}$ alone from our cross-correlation requires
external estimates of the \HI\ bias and stochasticity.  The linear bias of \HI\
is expected to be $\sim0.65$ to $\sim1$ at these redshifts
\cite{2010ApJ...718..972M, 2011MNRAS.415.2580K}.  Simulations to interpret
\citet{2010Natur.466..463C} find values for $r$ between $0.9$ and $0.95$
\cite{2011MNRAS.415.2580K}, albeit for a different optical galaxy population.
Measurements of the correlation coefficient between WiggleZ galaxies and the total
matter field are consistent with unity in this $k$-range (with $r_{m,{\rm opt}}
\gtrsim 0.8$) \cite{2011MNRAS.415.2876B}.  These suggest that our
cross-correlation can be interpreted as $\Omega_{\rHI}$ between $0.45\times
10^{-3}$ and $0.75 \times 10^{-3}$.

Measurements with Sloan Digital Sky Survey
\citep{2009ApJ...696.1543P} suggest that before $z=2$,
$\Omega_{\rHI}$ may have already reached $\sim0.4\times10^{-3}$. At low
redshift, 21\,cm measurements give $\Omega_{\rHI}(z\sim0)=(0.43\pm0.03)
\times10^{-3}$ \cite{2010ApJ...723.1359M}.  Intermediate redshifts are more
difficult to measure, and estimates based on Mg-II lines in DLA systems
observed with Hubble Space Telescope find $\Omega_{\rHI}(z\sim1)\approx(0.97\pm0.36)\times10^{-3}$
\citep{2006ApJ...636..610R}, in rough agreement with $z\approx0.2$
DLA measurements \citep{2011ApJ...732...35M} and 21\,cm stacking
\citep{2007MNRAS.376.1357L}.  This is in some tension with a model where
$\Omega_{\rHI}$ falls monotonically from the era of maximum star formation
rate \citep{2012MNRAS.420.2799D}.  Under the assumption that $b_{\rHI}=0.8,
r=1$, the cross-correlation measurement here suggests
$\Omega_{\rHI}\sim0.5\times 10^{-3}$, in better agreement,
but clearly better measurements of
$b_{\rHI}$ and $r$ are needed.  Redshift space
distortions can be exploited to break the degeneracy between $\Omega_{\rHI}$
and bias to measure these quantities independently of simulations
\cite{2008arXiv0804.1624W, 2010PhRvD..81j3527M}.  This will be the subject of
future work.

Our measurement is limited by both the number of galaxies in the WiggleZ fields
and by the noise in our radio observations. Simulations indicate that the variance
observed in our radio maps after foreground subtraction is roughly consistent
with the expected levels from thermal noise.  This is perhaps not surprising,
our survey being relatively wide and shallow compared to an optimal
LSS survey, however, this is nonetheless encouraging. 

\acknowledgements

We thank John Ford, Anish Roshi and the rest of
the GBT staff for their support;  Paul Demorest and Willem van-Straten for help
with pulsar instruments and calibration; and K.~Vanderlinde for helpful conversations.

K.W.M. is supported by NSERC Canada.
E.R.S. acknowledges support by NSF Physics Frontier Center grant PHY-0114422 to
the Kavli Institute of Cosmological Physics.
J.B.P. and T.C.V. acknowledge support under NSF grant AST-1009615.
X.C. acknowledges the Ministry of Science and Technology Project 863 (under grant
2012AA121701); the John Templeton Foundation and NAOC Beyond the Horizon
program; the NSFC grant 11073024. 
A.N. acknowledges financial support from the Bruce and Astrid McWilliams Center
for Cosmology.

Computations were performed on the GPC supercomputer at the SciNet HPC
Consortium.






\bibliography{main}

\end{document}